# Thaddäus Derfflinger's sunspot observations during 1802-1824: A primary reference to understand the Dalton Minimum

Hisashi Hayakawa (1-2)*, Bruno P. Besser (3-4), Tomoya Iju (5), Rainer Arlt (6), Shoma Uneme (7), Shinsuke Imada (7), Philippe-A. Bourdin (4-3), Amand Kraml (8)

(1) Graduate School of Letters, Osaka University, 5600043, Toyonaka, Japan (JSPS Research Fellow).
(2) UK Solar System Data Centre, Space Physics and Operations Division, RAL Space, Science and Technology Facilities Council, Rutherford Appleton Laboratory, Harwell Oxford, Didcot, Oxfordshire, OX11 0QX, UK
(3) Space Research Institute, Austrian Academy of Sciences, Graz, 8042, Austria
(4) Institute of Physics, University of Graz, Universitätsplatz 5/II, 8010 Graz, Austria
(5) National Astronomical Observatory of Japan, 1818588, Mitaka, Japan.
(6) Leibniz-Institut für Astrophysik Potsdam (AIP), An der Sternwarte 16, D-14482 Potsdam, Germany
(7) Institute for Space-Earth Environmental Research, Nagoya University, 4648601, Nagoya, Japan
(8) Sternwarte, Stift Kremsmünster 4550, Kremsmünster, Austria

* hayakawa@kwasan.kyoto-u.ac.jp/hisashi.hayakawa@stfc.ac.uk

**Abstract**

As we are heading towards the next solar cycle, presumably with a relatively small amplitude, it is of significant interest to reconstruct and describe the past grand minima on the basis of actual observations of the time. The Dalton Minimum is often considered one of the grand minima captured in the coverage of telescopic observations. Nevertheless, the reconstructions of the sunspot group number vary significantly, and the existing butterfly diagrams have a large data gap during the period. This is partially because most long-term observations have remained unexplored in historical archives. Therefore, to improve our understanding on the Dalton Minimum, we have located two series of Thaddäus Derfflinger's observational records (a summary manuscript and logbooks) as well as his Brander's 5.5-feet azimuthal-quadrant preserved in the Kremsmünster Observatory. We have revised the existing Derfflinger's sunspot group number with Waldmeier classification and eliminated all the existing 'spotless days' to remove contaminations from solar meridian observations. We have reconstructed the butterfly diagram on the basis of his observations and illustrated sunspot distributions in both solar hemispheres. Our article aims to revise the trend of Derfflinger's sunspot group number and to bridge a data gap of the existing butterfly diagrams around the Dalton Minimum. Our results confirm that the Dalton Minimum is significantly different from the Maunder Minimum, both in terms of cycle amplitudes and sunspot distributions. Therefore, the Dalton Minimum is more likely a secular minimum in the long-term solar activity, while further investigations for the observations at that time are required.

**Introduction:**

It is important to investigate and reconstruct solar activity of the past as it provides fundamental input for several fields such as the solar dynamo theory (Charbonneau, 2010; Atlt and Weiss, 2014; Auguston *et al*., 2015; Hotta *et al*., 2016), the solar-terrestrial relationship (Lockwood, 2013; Hayakawa *et al*., 2018d, 2019b), space weather (Cliver and Dietrich, 2013; Hayakawa *et al*., 2017, 2018c, 2019a; Toriumi *et al*., 2017, 2019), space climate (Hathaway and Wilson, 2004; Owens *et al*., 2011; Barnard *et al*., 2011; Usoskin *et al*., 2015; Hayakawa et al., 2019c; Pevtsov *et al.* 2019), terrestrial climate change (Gray *et al*., 2010; Lockwood, 2012; Owens *et al*., 2017), and for predictions of upcoming solar cycles (Svalgaard *et al*., 2005; Petrovay, 2010; Iijima *et al*., 2017; Upton and Hathaway, 2018). Excluding the solar cycle of approximately 11 years, solar activity has longer-term variations such as grand minima and grand maxima (Solanki *et al*., 2004; Solanki and Krivova, 2004; Usoskin *et al*., 2007; Clette *et al*., 2014; Inceoglu *et al*., 2015; Muscheler *et al*., 2016; Usoskin, 2017).

Therefore, it is important to investigate the properties of the sunspot number during the grand minima, on the basis of contemporary observations. Certain predictions suggest the possibility of another grand minimum in the near future (*e.g*., Lockwood, 2010; Barnard *et al*., 2011; Solanki and Krivova, 2011; Iijima *et al*., 2017; Upton and Hathaway, 2018). However, we have only two grand minima within the coverage of direct sunspot observations recorded using telescopes for about 400 years (Hoyt and Schatten, 1998; Clette *et al*., 2014; Clette and Lefevre, 2016; Vaquero *et al*., 2016; Svalgaard and Schatten, 2016), while grand minima and grand maxima are reported in millennial time scale compiled by multiple cosmogenic isotopes from tree-rings and ice cores (Solanki *et al*., 2004; Usoskin *et al*., 2007; Inceoglu *et al*., 2015; Usoskin, 2017; Wu *et al*., 2018).

Further, recent revisions of sunspot number based on historical documents require us to re-evaluate the solar activity for a longer time span (Clette and Lefevre, 2016; Vaquero *et al*., 2016). Reconsideration of historical documents also suggests that we should seek to new records (Vaquero

*et al*., 2007, 2011; Arlt, 2008, 2009, 2018; Hayakawa *et al*., 2018a, 2018b; Carrasco *et al*., 2015, 2016, 2018, 2019b; Denig and McVaugh, 2017), remove apparent continuous spotless days (Vaquero, 2007; Vaquero *et al*., 2016), and revise observations based on modern viewpoints (Vaquero *et al*., 2016; Svalgaard, 2017; Hayakawa *et al*., 2018d; Arlt *et al*., 2013, 2016; Fujiyama *et al*., 2019; Karoff *et al*., 2019; Jørgensen *et al*., 2019; Pevtsov *et al*. 2019). Based on these revisions, a long-term variation of solar activity was evaluated utilising multiple methodologies (Svalgaard and Schatten, 2016; Usoskin *et al*., 2016; Clette and Lefevre, 2016; Chatzistergos *et al*., 2017). The solar activity during the Maunder Minimum (*c.a.*, 1645–1715) was also reconsidered (Vaquero *et al*., 2015; Usoskin *et al*., 2015, 2017) and the scenario of its onset was notably rewritten (Vaquero *et al*., 2011).

In this context, it is discussed if the Dalton Minimum (*c.a.*, 1797–1827) should be considered as one of the grand minima (*e.g.*, Kataoka *et al*., 2012; McCracken and Beer, 2014) or one of the secular minima in the long-term solar activity (*e.g.*, Usoskin *et al*., 2015). So far, this "minimum" has been studied, including its amplitude and cyclicity (Schüssler *et al*., 1997; Sokoloff, 2004; Usoskin *et al*., 2007; Petrovay, 2010; Usoskin, 2017). After Wolf (1894), the amplitude and cycles of its primary part have been discussed with contemporary sunspot observations (Hoyt and Schatten, 1992a, 1992b), and in the auroral reports in Europe (Schröder *et al*., 2004). Recent studies provide some insights upon the discussions on its onset (Usoskin *et al*., 2009; Zolotova *et al*., 2011) with a revision of the sunspot number (Vaquero *et al*., 2016; Hayakawa *et al*., 2018a) and reconstructions of proxies of cosmogenic isotopes (Karoff *et al*., 2015; Owens *et al*., 2015). Further, the recovery of sunspot observations for this period is ongoing, for example, in the observations recorded by Jonathan Fisher during 1816–1817 (Denig and McVaugh, 2017) or by Franz Hallaschka during 1814–1816 (Carrasco *et al*., 2018), to improve the reconstruction of sunspot activity. These studies have demonstrated that the Dalton Minimum was probably considerably different from the Maunder Minimum in terms of the duration and the amplitude of solar cycles (*e.g.*, Miyahara *et al*., 2004; Usoskin *et al*., 2007, 2015; Vaquero *et al*., 2015)

Thaddäus Derfflinger was one of the most active and important long-term observers during the Dalton Minimum (see Figure 18 of Clette *et al*., 2014; Figure 2 of Svalgaard and Schatten, 2016; Figure 1 of Willamo *et al*., 2017). His sunspot observations were studied by Wolf (1894) long after Derfflinger's death and were adopted by Hoyt and Schatten (1998) as they were. However, Wolf explicitly mentioned that he did not consult the original manuscript but received the information through a letter from Schwab, one of Derfflinger's successors as the director of the Kremsmünster Observatory (Wolf, 1894, pp. 97-98). Further, the classification method of the sunspot groups seems slightly different from the early modern times to modern time and hence, is subjected to reconsideration (*e.g.*, Svalgaard, 2017). Therefore, in this study, we referred to the original manuscript in the Kremsmünster Observatory, re-examined Derfflinger's sunspot observations and to reconstruct the time series of the sunspot group number and measure the sunspot positions according to the records in the original manuscripts.

**2. Observers: Thaddäus Derfflinger and his Assistants**

Thaddäus Derfflinger (Figure 1) was born on 19 December 1748 at Mühlwang near Gmunden and passed away on 18 April 1824 in Kremsmünster (Fellöcker, 1864, pp. 91 – 159). He studied theology and mathematics at the University of Salzburg and received his priesthood ordination in Passau. Around 1776, he studied astronomy under Placidus Fixlmillner (1721–1791), the first astronomer of the Kremsmünster Observatory (N48°03′, E14°08′). When Fixlmillner passed away in 1791, Derfflinger took over the position of director of the observatory and remained there for 33 years until his death (Fellöcker, 1864, p. 92). Kremsmünster Observatory is not situated in Germany as reported in Hoyt and Schatten (1998) but in Austria under the rule of the Habsburg Empire (Fellöcker, 1864).

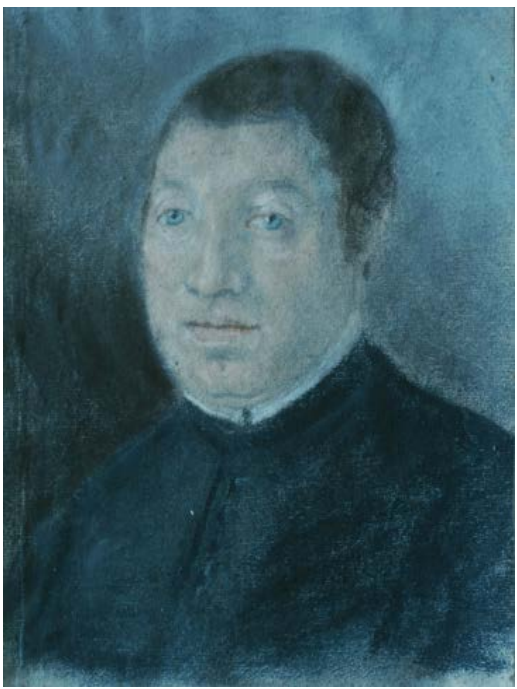

Figure 1: Derfflinger's portrait in a chalk drawing by Grinzenberger dated 25 April 1797 (courtesy: Kremsmünster Observatory).

While Derfflinger experienced the turmoil during the French invasion under Napoleon in 1800 and 1804–1805 (Fellöcker, 1864, p. 96), he continued his sunspot observations even during the second invasion. He was in regular contact with observatories in Vienna and Prague, including with the contemporary sunspot observer Franz Hallaschka (Fellöcker, 1864, p. 99–100) whose sunspot records have been recently recovered (Carrasco *et al*., 2018). During the last two decades of Derfflinger life, he suffered from the degradation of his eyesight and lost his left eyesight in spring 1819 (Fellöcker, 1864, p. 92). The loss is partially because of his long-term sunspot observations. However, he maintained his right eyesight and supervised the sunspot observations until 21 March 1824, one month before his death (Fellöcker, 1864, pp. 93 – 111).

Within the monastery, he had two assistants: Benno Waller (1758–1833) and Leander Öttl (1757–1849), two other monks of the confraternity of Benedictines. He had at least three more assistants outside of the monastery: Johann Illinger (1724–1800), Simon Lettenmayr (father: 1757–1834), and Simon Lettenmayr (son: 1787–1868). Johann Illinger had worked in the observatory since the time of Fixlmillner. Simon Lettenmayr (father) worked not only on the construction and reparation of the monastery buildings but also as an observational assistant. His son, also named Simon Lettenmayr, accompanied Derfflinger during his visit to Prague in 1816 and was advised by Hallaschka to improve and construct observational instruments. He also received basic education in meteorology and astronomy from the teachers of Kremsmünster School and performed magnetic observations under the supervision of the observatory directors: Derfflinger, Bonifaz Schwarzenbrunner (1790–1830), and Marian Koller (1792–1866). Eventually, his eyesight considerably degraded and he retired (Fellöcker, 1864, pp. 111–112).

**3. Observational Records:**

Derfflinger's sunspot records are currently preserved in the directorate archives of the Kremsmünster Observatory. The sunspot drawings have been recorded both in the meteorological logbooks (v. 2–5) and the summary manuscript entitled 'Overview of the sunspots which were observed on the observatory of Kremsmünster since 26 September 1802 until 1824 inclusive; then as of 26 July 1848 (*Uibersicht der Sonnenmackeln welche auf der Sternwarte zu Kremsmünster seit dem 26. September 1802 beobachtet wurden, bis 1824 inclusive; dann vom 26. Juli 1848*)' (see Appendix 1). It is inferred that there may have been original daily sunspot drawings besides these manuscripts, made directly at the telescope, as sunspot drawings are cruder in the meteorological logs and more detailed in the summary manuscript.

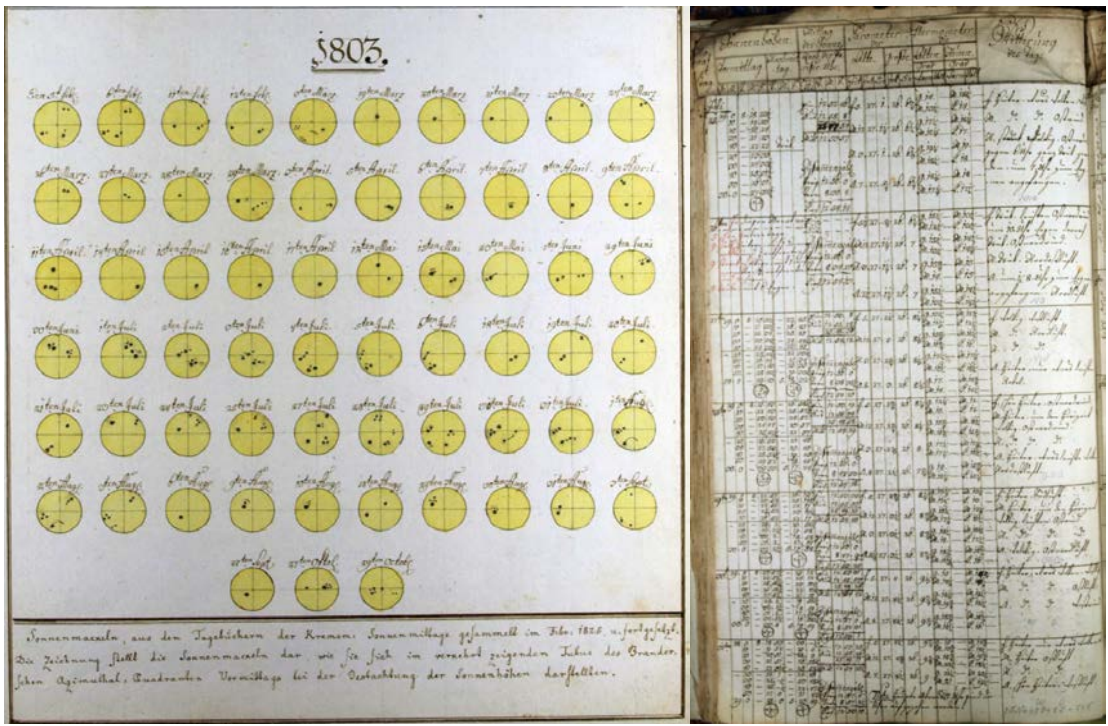

Figure 2: (a) Sunspot observations in 1803 in Derfflinger's summary manuscript with its footnote (p. 1) and (b) Sunspot drawings dated 25–31 July 1803 involved in meteorological logbooks (v. 2, p. 122). In the right panel, sunspot drawings are shown as tiny circles including dots in which are placed in a table.

The summary manuscript (Figure 2(a)) explains in its footnote when and why it was compiled: 'Sunspots, from the diaries of Sun noon observations in Kremsmünster, summarized in February 1825 and following years. The drawings illustrate the sunspots like they have been depicted with the inverting lens tube of the azimuthal quadrant of Brander ante meridiem at the observation of the solar altitude' (Figure 2). This footnote shows that the sunspot observations were a by-product of the regular solar elevation observations to determine local solar noon since 1802.

Accordingly, this summary collected sunspot drawings around mid-day and was compiled in February 1825, soon after the death of Derfflinger. It seems to have been planned to collect further sunspot drawings after 1825, as shown with the unfilled margin for 1825 without sunspot drawings (summary manuscript, p. 7). Further, a sunspot drawing on 26 July 1848 with at least 9 sunspot groups was included by an anonymous observer, possibly associated with Augustin Reslhuber, director of Kremsmünster Observatory at the time. On this date, both Schwabe and Shea reported 6 sunspot groups (Vaquero *et al*., 2016).

**4. Instruments and Telescopes:**

Since 1802, the Kremsmünster Observatory (Figure 3(a)) monitored the position of the Sun on a regular basis to time the true local noon with the aid of a transportable quadrant. By measuring various timings of given elevations of the Sun in the morning and in the afternoon, the solar culmination can be computed and the time of true local noon determined. During these measurements, sunspots have been recognised quite frequently and small sunspot sketches have been recorded into the meteorological logbooks (Figure 2(b)).

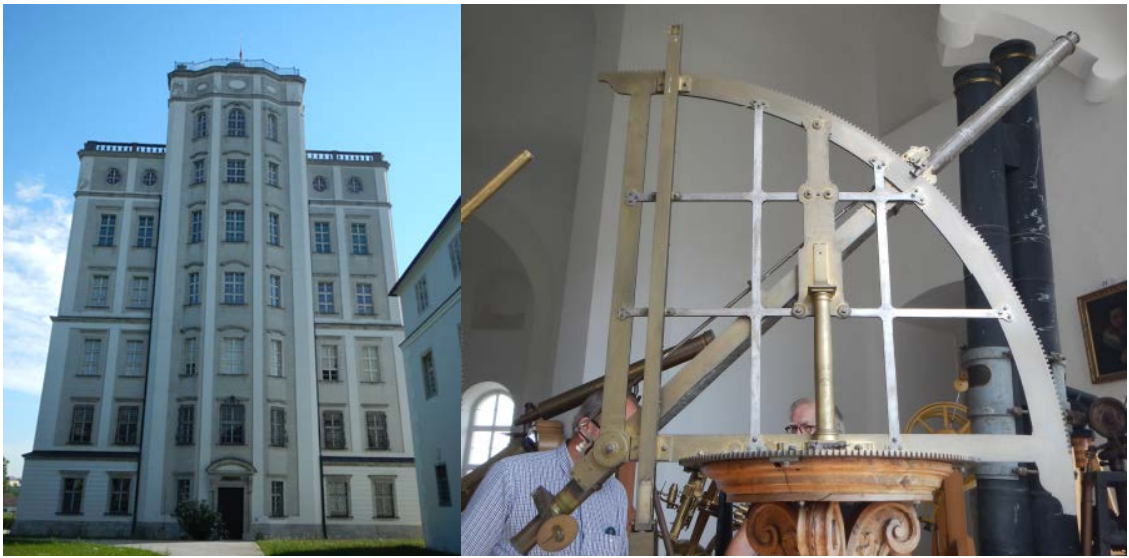

Figure 3: (a) The Kremsmünster Observatory (N48°03′; E14°08′); (b) Brander's 5.5-feet azimuthal-quadrant preserved in the Kremsmünster Observatory.

Several quadrants were used at this observatory to determine local noon (off the meridian). In principle, one of the two mural quadrants mounted in the observation hall (6th and 7th floor), the one to the south (B), as depicted in a drawing of Fellöcker (1864) seems to have been used. Among them, Derfflinger seemed to use Brander's 5.5-feet azimuthal-quadrant (Wolf, 1894, p. 98) with a micrometre manufactured in Paris (Fellöcker, 1864, p. 60; Figure 3(b)). This instrument has a height of approximately 266 cm, with a telescope with a focal length of approximately 174 cm, the radius of the quarter angular arc is 95 cm, and is mounted on an oaken stand. The instrument was in regular use until 1824 (Fellöcker, 1864, p. 12, footnote 10), on the basis of contemporary length unit in Austria (Aldefeld, 1838). The summary manuscript (Figure 2(a)) records that this quadrant had an inverting lens, thus, indicating it to be a Keplerian telescope.

## 5. Sunspot Groups Recorded in Derfflinger's Manuscript:

Examining the summary manuscript and meteorological logbooks, we identified sunspot observations for 487 days. We have counted their group number with the Waldmeier classification (Kiepenheuer, 1953) and summarized our result[1]. The total number is notably less than the number of days with sunspot observation (789 days) in the existing dataset (Hoyt and Schatten, 1998; Vaquero *et al.*, 2016). This is mainly because we observed that the existing spotless days are likely to be solar elevation observation without sunspot drawings. We eliminated these data.

Wolf (1894, pp. 98-99) had cited Franz Schwab to have stated that, 'On those days when the sun was observed, without any sunspots being noted, the symbol ○ was applied, which does not mean that the sun was spotless, but it does indicate that the observer did not notice anything of particular interest on the solar surface; however, sometimes the sketch could have been omitted for lack of time'. Nevertheless, Wolf (1894, pp. 99–103) misleadingly substituted the ○ symbol by the slimmer 0 to fit it into his published tables. These '0's have been incorporated to Hoyt and Schatten (1998) as spotless days.

As mentioned earlier, these sunspot observations are by-products of meridian observations. Analogous studies on observations of corresponding solar elevations showed that it is not straightforward to reconstruct solar activity from the solar meridian observations, as shown in the meridian solar observations by the Royal Observatory of the Spanish Navy (Vaquero and Gallego, 2014). Similarly, the meridian solar observations in Bologna (Manfredi, 1736) and by Hevelius (1679) had been misinterpreted as spotless days (Vaquero, 2007; Clette *et al.*, 2014; Carrasco *et al.*, 2016).

(a)

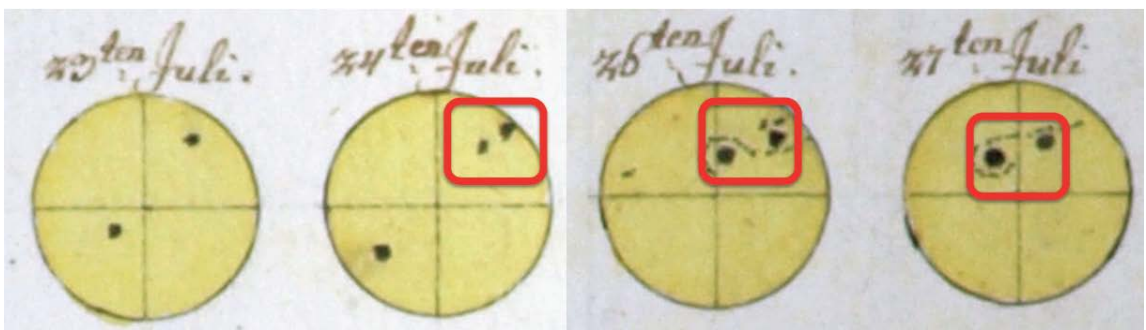

---

[1] https://www.kwasan.kyoto-u.ac.jp/~hayakawa/data

(b)

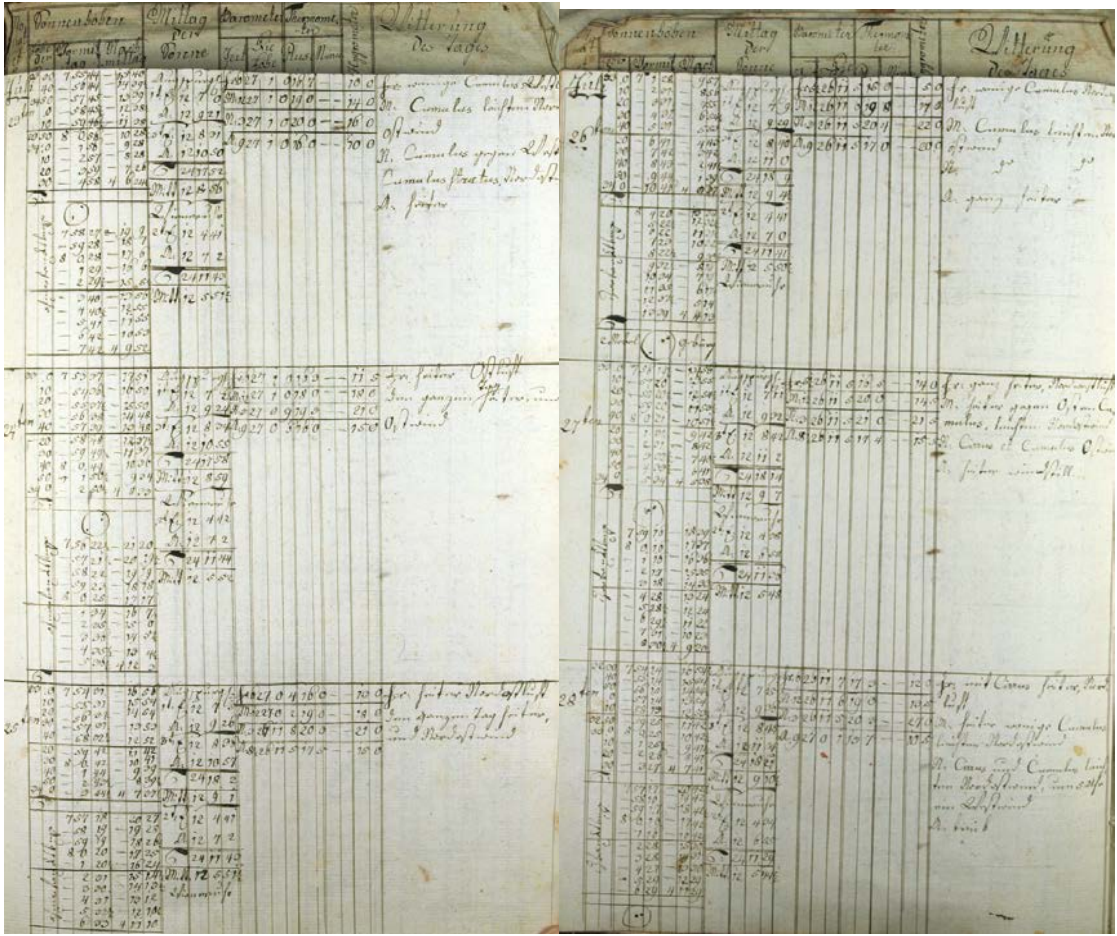

Figure 4: Consequence of sunspot observations on 23–27 July 1818 in (a) the summary manuscript (p. 5) and (b) the logbook (v. 4, pp. 435–436).

In certain instances, the meteorological logbooks show the meridian observations without sunspot drawings while subsequent solar elevation observations show multiple sunspots as observed in Figure 4. Here, while the absence of sunspot drawing on 25 July 1818 had been considered as a spotless day in the existing database (Hoyt and Schatten, 1998; Vaquero *et al*., 2016), this seems unlikely given the fact that a sunspot group (red box in Figure 4(a)) moved from the eastern limb to the disc centre and disappeared only on 25 July 1818. Therefore, it is highly unlikely that the sunspot groups disappeared in between for a day within a sequence of observations, while their group numbers and relative locations are almost similar on other observing days. Furthermore, even for the dates of solar meridian observations without sunspot drawings, the summary manuscript occasionally recorded sunspot drawings (*e.g*., 15 and 31 July 1816). Therefore, we have concluded

Hayakawa et al. (2020) Thaddäus Derfflinger's sunspot observations during 1802-1824, *The Astrophysical Journal*, doi: 10.3847/1538-4357/ab65c9

that Derfflinger's solar elevation observations without sunspot drawings should not be considered as spotless days, but as an absence of observational data.

Excluding the apparent spotless days, we have also revised 7 observational dates, eliminated 8 observational dates, and added 7 observational dates against the register in the existing dataset (Hoyt and Schatten, 1998; Vaquero *et al*., 2016), as per the summary sheet and original meteorological logbooks. The summary manuscript records a sunspot drawing on 8 August 1824, whereas the logbook (v. 5, p. 342) does not show any solar elevation observation on that date. This might be because Derfflinger or one of his subordinates may have observed the solar disk on this date outside of the observatory. This incomparable instance indicates that there should have been original drawings made at the telescope from which both the summary sheets and the logbook sketches were copied. Further investigations in the Kremsmünster archives are required in this regard.

Figure 5 shows a time series of revised sunspot group numbers of Derfflinger contextualized on the existing database for the sunspot group number (Vaquero *et al*., 2016) and additional sunspot observations by Fisher (Denig and McVaugh, 2017) and Hallaschka (Carrasco *et al*., 2018).

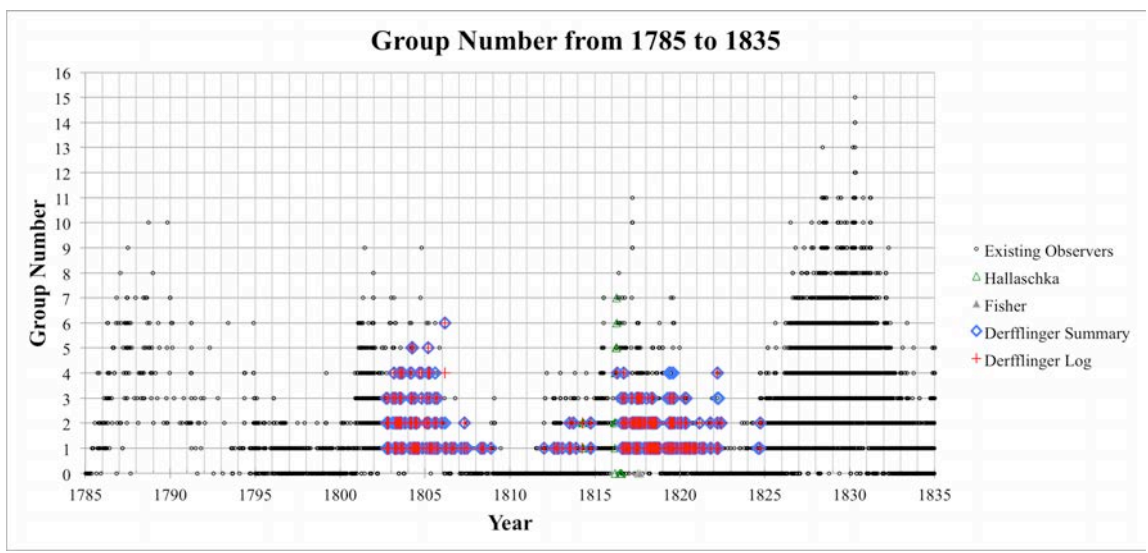

Figure 5: Revised sunspot group number of Derfflinger (blue diamond in the summary manuscript and red cross in the logbooks) contextualized upon the sunspot group number of the existing observers (black dot: Vaquero *et al*., 2016) and additional sunspot observations by Fisher (gray triangle: Denig and McVaugh, 2017) and Hallaschka (green triangle: Carrasco *et al*., 2018).

In Figure 5, we have incorporated the sunspot group number in Derfflinger's summary manuscript and logbook as they are and contextualised them upon the existing sunspot group number by contemporary observers. The summary manuscript and logbooks occasionally provide observations with different group number and observations on different days.

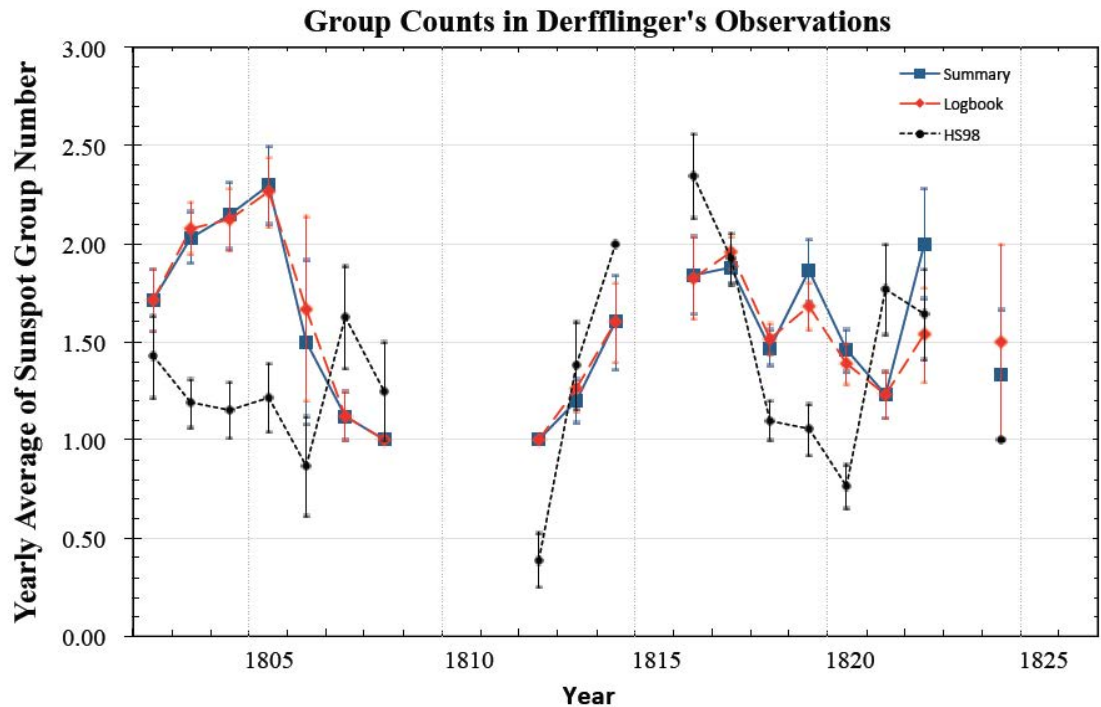

Figure 6: Derfflinger's revised yearly sunspot group number in the summary manuscript (red diamond) and logbook (blue square) with standard errors from averaging, in comparison with the yearly number before our revision (black circle; Hoyt and Schatten, 1998; Vaquero *et al*., 2016).

As observed in figure 6, based on the revised sunspot group number series, we have revised Derfflinger's yearly sunspot group number during 1802–1824 in comparison with the number before our revision in the existing datasets (Hoyt and Schatten, 1998; Vaquero *et al*., 2016). Our revision shows a significantly different trend of yearly Derfflinger's sunspot group number (Figure 6) in comparison with that before our revision (Figure 7). We observed that Derfflinger's trend is much more consistent with the Sunspot Number (Version 2; Clette *et al*., 2014; Clette and Lefevre, 2016) than the group sunspot number series (Svalgaard and Schatten, 2016; Usoskin *et al*., 2016) in Cycle 5. However, it is more consistent with the group sunspot number series than the Sunspot Number (Version 2; Clette *et al*., 2014; Clette and Lefevre, 2016) in Cycle 6. The cycle amplitude seems slightly larger in Cycle 5 than in Cycle 6.

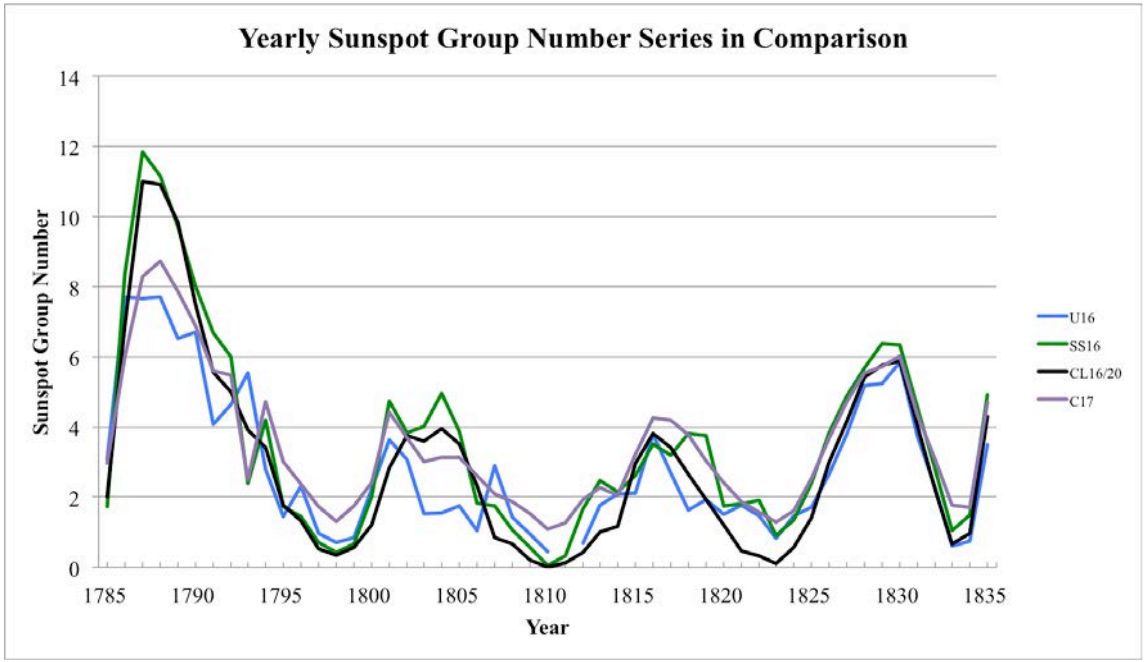

Figure 7: Comparison of the yearly international sunspot number (Clette and Lefevre, 2016) divided by 20 (Muñoz-Jaramillo and Vaquero, 2018) and existing yearly sunspot group number series with backbone method (green curve = Svalgaard and Scatten, 2016; purple curve = Chatzistergos *et al.*, 2017) and active day fraction method (blue curve = Usoskin *et al.*, 2016).

**6. Measurements of the Sunspot Positions**

It is difficult to determine the heliographic coordinates of the sunspots from Derfflinger's drawings as he had not recorded an explicit time for each drawing. There are horizontal and vertical lines in the drawings, which are supposed to be parallel to the horizon and pointing to the zenith, respectively, since the manuscripts mention that the observations are made with an azimuthal quadrant. The manuscript also states that the images are upside-down, *i.e.* the lower end of the vertical line points to the zenith. An indication of the observing time comes from the logbooks in which sketches of the sunspot drawings are inserted above, below, or within the solar elevation timings.

  Hence, we are using two ways of fixing the position angle of the solar disk to obtain the heliographic coordinates. The first method uses the nearest time of the solar elevation measurements to fix the position angle of the Sun, as it appeared at that time in the sky in a horizontal coordinate system while using an ephemeris provided by the JPL Horizons system[2]. These times give subjectively reasonable results for the majority of days. The second method can be employed if the spot(s) were drawn on several days in a row. Assuming the heliographic positions have not changed over the course of the days, the position angles are obtained along with the longitudes and latitudes using Bayesian inference (Arlt *et al.*, 2013). This method is called rotational matching. By utilising the elevation times, this method was used when a sequence of days with the same spots did not show a consistent progression of the spots. If in such a sequence, only a single day contained an outlier, we adapted the time of observation to a moment when the position angle results in a reasonable progression of the spots. These manually found times typically fall later in the day and indicate that some observations may not have been made in direct connection with the elevation measurements.

We also employed a correction to the clocks, as the solar elevation measurements provide us with the local solar time, *i.e.* the meridian passage of the Sun. When compared with the solar equation of time, we obtained a correction to the Kremsmünster clocks. The maximum clock correction applied reached −1.22 h on 13 November 1804, after 20 days of bad weather, when the clocks had not been adjusted according to the solar elevation measurements. After the solar minimum starting Cycle 6, the clocks were more precise, and very few observations show deviations of more than 15 minutes.

The butterfly diagram resulting from 2210 sunspot positions of 487 observations[3] is shown in Figure 8. The spot locations during Cycle 5 show a migration of activity towards the equator. There are several spots on the equator, a phenomenon which may be attributed to the low accuracy of the drawings and that we are not plotting group centres (*e.g.*, Figure 9 of Hathaway, 2015), but all are individual spot locations. As the groups are apparently plotted in a magnified manner, the individual spots can easily populate the equator, even though the group centre is clearly in one of the hemispheres. Cycle 6 looks similar to Cycle 5, while the observational gaps in 1814 and 1815 hide features of the early phase of the cycle. Nevertheless, our butterfly diagram shows the asymmetric appearance of high-latitude spots in the beginning phase of Cycle 6. Figure 8 shows that spots in the northern hemisphere appeared in late 1811, whereas those in the southern hemisphere appeared in mid-1813. Similarly, some high-latitude spots indicate the beginning of Cycle 7 in 1822, whereas from sunspot numbers alone, the cycle minimum is usually placed in early 1823 (Hathaway, 2015). Cycle 7 appears to start with spots in the northern hemisphere, but their total number is small. In summary, we conclude that the butterfly diagram is—while slightly asymmetric—generally compatible with its modern shape and that the differences to modern graphs are not significant and attributable to the limited accuracy of the observations.

---

[2] https://ssd.jpl.nasa.gov/horizons.cgi
[3] https://www.kwasan.kyoto-u.ac.jp/~hayakawa/data

Error margins of the heliographic positions were obtained in the following way. As the primary unknown quantity in the analysis is the position angle of the solar disk, we assumed a general uncertainty of ±10 degrees in the position angle. We then employed this uncertainty to the conversion of the spot locations into heliographic positions and obtained individual uncertainties for the heliographic longitudes and latitudes. In the case of rotational matching, the results are probability density distributions of all unknown quantities with 68% confidence intervals. We did not measure nearly circular arcs around sunspots that appear to denote penumbrae of evolving spots. However, non-circular arcs rather look like chains of smaller spots and were measured.

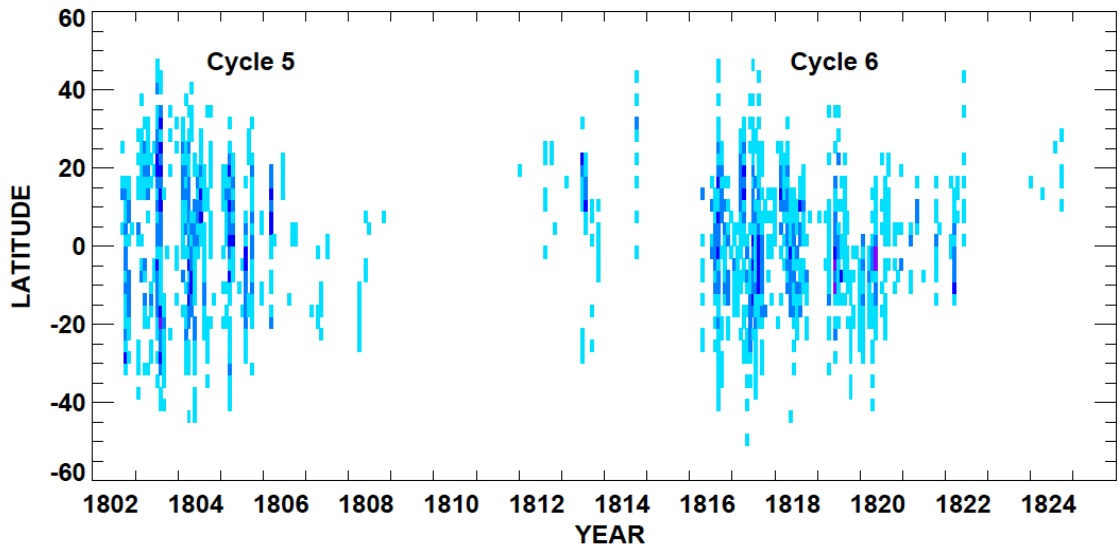

Figure 8: A reconstructed butterfly diagram for sunspot positions in Derfflinger's manuscript. This diagram is divided into blocks of 27-day duration at 3 degrees latitude range. The numbers of spots falling into these bins are counted. The darkness of the blue represents the number of spots counted in a bin. The spots with light blue represent 1 and 2 spots, those with medium blue represent 3–5 spots, those with dark blue represent 6–8 spots, and those with violet represent 9–12 spots.

**10. Conclusions and Outlooks**

In this article, we have examined Derfflinger's sunspot observations on the basis of his original records. Derfflinger's observations are currently preserved in the directorate archives of the Kremsmünster Observatory as a summary manuscript and meteorological logbooks (v. 2–5). Derfflinger conducted his sunspot observations from 1802 to 1824 with aids of his assistants, as by-products of solar elevation observations. Derfflinger used Brander's 5.5-feet azimuthal-quadrant for the observations of sunspots and of the solar elevation.

Examining his original observational records, we have discovered that the existing 'spotless days' were contaminations from solar elevation observations without sunspot drawings. Accordingly, these 'spotless days' were eliminated, as they do not necessarily mean the absence of sunspots. In contrast, we found evidence that there were sunspots at least in some of those removed data. We have also revised 7 observational dates, eliminated 8 observational dates, and added 7 observational dates against the register in the existing dataset (Hoyt and Schatten, 1998; Vaquero *et al.*, 2016). We then applied the Waldmeier classification to revise Derfflinger's group sunspot number. The revised Derfflinger's trend shows that the amplitude of Cycle 5 is slightly higher than that of Cycle 6. The revised trend seems rather consistent with the Sunspot Number (version 2) in Cycle 5, whereas his revised trend in Cycle 6 is more consistent with the group sunspot number series.

We have reconstructed the butterfly diagram on the basis of Derfflinger's sunspot observations and have filled the existing data gap (see Muñoz-Jaramillo and Vaquero, 2019). The reconstructed butterfly diagram demonstrates no significant asymmetry of sunspot distributions like that of the Maunder Minimum (Ribes and Nesme-Ribes, 1993). We observed considerable sunspots near the solar equator, probably due to the limited quality of Derfflinger's sunspot observations.

Our revision shows that Derfflinger's sunspot cycles during the Dalton minimum have a slightly higher amplitude of the solar activity than previously considered. The data gap of the butterfly diagram in this period has been filled and does not show extremely asymmetric sunspot distributions like those of the Maunder Minimum. Our reconstruction shows that the Dalton minimum was significantly different from the Maunder minimum, either in terms of cycle amplitude (*c.f.*, Usoskin *et al.*, 2015), its duration (*c.f.*, Miyahara *et al.*, 2004; Vaquero *et al.*, 2015), or its more symmetric butterfly diagram (*c.f.*, Ribes and Nesme-Ribes, 1993).

Primarily, the reconstructed cycles during the Dalton Minimum were approximately 11 years (≈ 12 years), while the period during the Maunder minimum was probably either considerably shorter (Vaquero *et al.*, 2015) or longer (Miyahara *et al.*, 2004) than 11 years. A peculiarity of the early Dalton minimum was a 'hiccup' in the cycle period. This may have been either a very long cycle of approximately 15 years duration (Hathaway, 2015) or a short cycle followed by a very weak one of, respectively, a little less than 9 years and more than 7 years (Usoskin *et al.*, 2009). That period falls before Derfflinger's observations.

Our study hints to an understanding of the Dalton minimum as not towards a grand minimum characterized with extremely weak or even collapsed solar dynamo cycles, but more towards a secular minimum in the long-term solar activity slightly longer cycles with low activity. This notion is more consistent with a solar dynamo that continued to produce a reasonable number of sunspots during the Dalton minimum. Potential deviations from the average cycle length are probably determined by quantities which are difficult to access in historical observations, such as the meridional circulation, stochastic variations in the convective patterns, or the internal rise time of magnetic flux to the solar surface (e.g., Charbonneau, 2013, Chapter 4; Fournier *et al*., 2018).

Nevertheless, it has been determined that Derfflinger did not record spotless days during his observations and the cycle amplitudes during the Dalton Minimum may be revised slightly downward. We are required to carefully revise the data of other contemporary sunspot observers during the Dalton minimum, revise the actual group number, and define the actual spotless days. Additionally, spot areas are to be studied further owing to their seemingly exaggerated size as observed in other historical observations using aerial imaging method (see *e.g*., Fujiyama *et al*., 2019; Karachik *et al*., 2019). Further research on the Dalton minimum will improve our knowledge of solar activity during the solar minima or the suppressed solar cycles.

**Acknowledgement**

We thank Kremsmünster Observatory for permitting access to Derfflinger's manuscripts and for the reproduction of some observational records as well as preserving these records. HH thanks F. Clette and S. Toriumi for their helpful comments. This research has been conducted with aids of KAKENHI Grant Number 15H05812 (PI: K. Kusano), JP15H05816 (PI: S. Yoden), and JP17J06954 (PI: H. Hayakawa), as well as the Austrian Science Foundation (FWF) project P 31088 (PI: U. Fölsche) and the Deutsche Forschungsgemeinschaft grant number AR355/12-1 (PI: R. Arlt). This work has been partly merited from participation to the International Space Science Institute (ISSI, Bern, Switzerland) via the International Team 417 "Recalibration of the Sunspot Number Series".

**Appendix 1: Historical Sources**

Summary Manuscript: *Uibersicht der Sonnenmackeln, welche auf der Sternwarte zu Kremsmünster Seit dem 26. September 1802 beobachtet wurden. bis 1824 inclusive; dann vom 26. Juli 1848*, MS, Direktions-Archiv der Sternwarte Kremsmünster

Logbook (v.2): *Meteorologische Beobachtungen zu Kremsmünster 1801-1807*, **II**. Bd. MS, Direktions-Archiv der Sternwarte Kremsmünster

Logbook (v.3): *Meteorologische Beobachtungen zu Kremsmünster 1808-1813*, **III**. Bd. MS, Direktions-Archiv der Sternwarte Kremsmünster

Hayakawa et al. (2020) Thaddäus Derfflinger's sunspot observations during 1802-1824, *The Astrophysical Journal*, doi: 10.3847/1538-4357/ab65c9

Logbook (v.4): *Meteorologische Beobachtungen zu Kremsmünster 1814-19*, **IV**. Bd. MS, Direktions-Archiv der Sternwarte Kremsmünster

Logbook (v.5): *Meteorologische Beobachtungen zu Kremsmünster 1820-25*, **V**. Bd. MS, Direktions-Archiv der Sternwarte Kremsmünster

Hayakawa et al. (2020) Thaddäus Derfflinger's sunspot observations during 1802-1824, *The Astrophysical Journal*, doi: 10.3847/1538-4357/ab65c9

Hayakawa et al. (2020) Thaddäus Derfflinger's sunspot observations during 1802-1824, *The Astrophysical Journal*, doi: 10.3847/1538-4357/ab65c9

Hayakawa et al. (2020) Thaddäus Derfflinger's sunspot observations during 1802-1824, *The Astrophysical Journal*, doi: 10.3847/1538-4357/ab65c9